\documentclass[prl,aps,reprint,letterpaper,nofootinbib,superscriptaddress,floatfix]{revtex4-2}
\usepackage{times}
\usepackage{graphicx}
\usepackage{epsfig}
\usepackage{amsmath,amssymb,amsthm}
\usepackage{dsfont}
\usepackage{color,xcolor}
\usepackage[normalem]{ulem}
\usepackage[normalem]{ulem}
\usepackage{caption}
\usepackage[colorlinks=true,linkcolor=blue,citecolor=blue,urlcolor=blue,breaklinks]{hyperref}
\usepackage{comment}
\usepackage{ragged2e}
\usepackage{braket}
\usepackage{physics}
\usepackage{subfig}
\usepackage{bbm}
\usepackage{bm}
\usepackage{dsfont}

\usepackage[T1]{fontenc}
\usepackage{braket}

\begin{document}

\title{Quantum coherence leveraged agnostic phase estimation}
\author{H. S. Karthik}
\email{hsk1729@gmail.com}
\affiliation{International Centre for Theory of Quantum Technologies (ICTQT), University of Gda{\' n}sk, 80-308 Gda{\' n}sk, Poland}

\begin{abstract}
Quantum metrology concerns improving the estimation of an unknown parameter using an optimal measurement scheme on the quantum system. More the optimality of the measurement, the better will be the improvement in sensing the value of the unknown parameter. Pertaining to the case of a two level system (qubit) undergoing rotation, a typical metrological task concerns the estimation of the angle of rotation ($\tau$) given the information about the axis of rotation ($\theta,\phi$). The method for garnering information about ($\tau)$ is through maximizing the Fisher information. In the absence of the axis-knowledge, the optimality of the metrological task reduces drastically. Drawing inspiration from recent works leveraging entanglement to connect closed time-like curves and metrology, we overcome this limitation (lack of the knowledge of the rotation axis) using an ancilla assisted protocol. Here, the probe and the ancilla interact through a coherently controlled superposition of unitary evolutions. The \emph{quantum coherence} in the initial state of the ancilla forms as the resource aiding the protocol. By measuring the ancilla in the same coherent basis in which it was prepared, we achieve optimal Fisher information about the rotation angle, independent of the axis parameters. Notably, this resource-efficient and operationally simple agnostic sensing alternative is independent of requiring entanglement in the initial joint state of the probe and the ancilla or entangling measurements, yet accounts for maximum Fisher information about the angle of rotation. 
\end{abstract}
\maketitle
\emph{Introduction}:--Metrology concerns with techniques related to improving precision of measurements with respect to estimating a parameter. The original formulation of a metrological task for classical systems relied on the principles of classical physics. However, with the advent of quantum theory, manipulating and utilizing the quantum nature of a system has lead to many critical developments in quantum metrology \cite{braunstein1994statistical,giovannetti2006quantum, paris2009quantum,toth2014quantum}. 

A paradigmatic task in quantum metrology is estimating an unknown phase (the terms phase and rotation angle are used interchangeably) encoded via a unitary transformation typically of the form $\rho_\tau = e^{i H \tau}~ \rho~  e^{-i H \tau}$, where $U =  e^{-i H \tau}$ represents the unitary transformation and $H$ is a known Hermitian operator that generates the phase $\tau$. Assuming a pure input state, $\rho = \ket{\psi}\bra{\psi}$, the pertinent goal here is the estimation of the phase $\tau$ by performing an appropriate measurement on $\rho_\tau$. Crucially, standard quantum metrology protocols presuppose full knowledge of the generator $H$, which determines the structure of the optimal probe state. For a qubit, this optimal probe state for phase estimation corresponds to a coherent superposition of the eigenstates, $\ket{a_i}$ (with the corresponding eigenvalue $a_i$), of the Hamiltonian generating the transformation, $H = \sum_i a_i \ket{a_i}\bra{a_i}$ thereby maximizing sensitivity to the phase shift  \cite{braunstein1994statistical,giovannetti2006quantum,giovannetti2011advances,toth2014quantum}. Furthermore, the optimality of the measurement depends on $H$, implying that any lack of knowledge about $H$ precludes optimal probe state preparation and measurement optimization. Consequently, conventional phase estimation protocols fail when $H$ is not characterized.

This limitation poses a significant challenge for quantum metrology to operate in scenarios where (characteristics of) the generating dynamics are only partially known or difficult to calibrate in advance, say, in fluctuating or time dependent magnetic fields. Overcoming this constraint requires novel protocols that do not rely on prior knowledge of the Hamiltonian, but instead leverage intrinsic quantum resources such as coherence or entanglement to enable robust phase estimation.

Recent advances have begun to address metrological scenarios where the generator $H$ is not fully known. In \cite{arvidsson2023nonclassical}, a protocol was introduced that enables phase estimation by retrospectively preparing the probe in an optimal state through entanglement with an ancilla and a measurement conditioned on the revealed unitary $U = e^{-i H \tau}$ and \cite{song2024agnostic} demonstrated an agnostic metrology protocol that involves the following steps: a probe-ancilla pair in a maximally entangled state, the probe undergoing an unknown unitary transformation and a final joint measurement in the original entangled basis. Both the approaches focus on utilizing the equivalence of entanglement manipulation and its connection to closed time-like curves (CTC)---hypothetical world lines that travel back in time \cite{godel1949example,svet11,MTY88,MT88,D91,LMG+11,LMGR+11, RBM+14}. Both these protocols mainly illustrate that entangled state preparation can overcome the limitation of the complete knowledge of the phase-generating Hamiltonian (see also \cite{song2025superconducting} for a very recent work).

In this work, we leverage \emph{quantum coherence} to present a metrological protocol in which one can optimally estimate the phase and be agnostic of the Hamiltonian generating it. Unlike previous approaches that rely on entanglement either in state preparation or in measurement, our protocol exploits coherence in the ancilla as the primary resource. To this end, we demonstrate that \textit{quantum coherence} alone can be harnessed in enabling optimal estimation of the phase, $\tau$. Specifically, when a coherent ancilla controls the application of the unitary $U$ and its inverse ($U^{-1}$), we show that our protocol can enable Hamiltonian-independent phase estimation. This ability to (create) control and erase which-operation information mirrors the investigations on delayed-choice experiments \cite{kaiser2012entanglement,lee2014experimental,ma2016delayed}.  

 In the single-qubit setting, conventional phase estimation fails without the knowledge of $H$ \cite{chapeau2016optimizing,song2024agnostic,altorio2016metrology}. In contrast, our protocol-enabled by leveraging the coherence of the ancilla controlling the transformations $U$ and $U^{-1}$  (generating the phase)- shows that the possibility of agnostic phase estimation can be realized. To contextualize this result, we compare the strategies and the performance of the phase estimation protocols involving the single qubit probe, coherently controlled transformation of the probe and that of the entangled probe-ancilla.

 Interestingly, in a recent work, \cite{chapeau2021quantum}, the author explored a parameter estimation protocol utilizing superposition of quantum channels. However, our proposed metrological protocol predominantly differs from \cite{chapeau2021quantum} on two main aspects:(a) we consider using the superposition of the unknown unitary and its inverse--an arrangement not previously analyzed in the context of phase estimation (b) we identify that this unique superposition offers the route to agnostic phase estimation. Additionally, our work also offers a distinctive interpretation in terms of time-symmetric or time-travel like dynamics arising from leveraging the coherence of the ancilla in controlling the transformations $U$ and $U^{-1}$, a perspective not previously explored in the context of phase estimation. 
 
 Furthermore, \cite{xia2024nanoradian} considered a protocol for rotation-angle estimation using indefinite-time-direction dynamics. Here, an \emph{effective} indefinite-time-direction evolution using polarization as a control system gives rise to a state-dependent superposition of forward  ($U(g)$) and backward evolution ($U^{\dag}(g)$) where $g$ refers to the parameter to be estimated. Notably, both the protocols (ours in comparison to that of  \cite{xia2024nanoradian}) use coherence as the main resource. However, our protocol is much more general in terms of being agnostic to the axis, whereas theirs focuses on maximizing precision for a \textit{known} axis using high-orbital angular momentum beams and novel dynamical control. 

\begin{figure}[ht!]
\centering
        \includegraphics[width=0.85\linewidth]{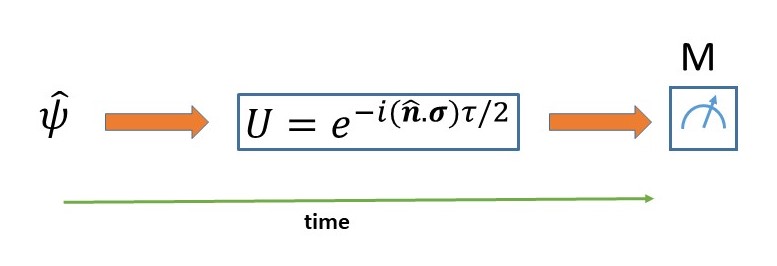}
   \caption{Single qubit sensing protocol--time runs from left to right}
   \label{fig:unitaryest}
\end{figure}

\emph{Preliminaries:--} We begin by reviewing the standard qubit phase estimation protocol, which requires prior knowledge of the Hamiltonian $H$ to achieve optimal performance. The phase $\tau$ is encoded via the unitary transformation $U = e^{-i H \tau}$ acting on an initial quantum probe state~(see Fig. \ref{fig:unitaryest}). The sensitivity of the resulting probability distribution to variations in $\tau$ is quantified using the classical Fisher Information (FI), defined as $F(\tau)=\sum_{m}\frac{(\partial_{\tau}p(m|\rho_\tau)^2}{p(m|\rho_\tau)}$, where $p(m|\rho_\tau)$ is the outcome probability of measurement $m$ on the evolved state  $\rho_\tau$. The variance in the estimator $\hat{\tau}$ is bounded from below by the \textit{ Cram\'er-Rao} (CR) bound,  $Var(\hat{\tau}) = 1/(N F(\tau))$, where $N$ is the number of statistical trials. The Quantum FI (QFI) $\mathcal{F}_Q(\tau|\rho_\tau)$ upper bounds the classical FI \cite{braunstein1994statistical}: $\mathcal{F}_Q(\tau|\rho_\tau) \geq F(\tau|\rho_\tau)$, implying the quantum version of the \textit{ Cram\'er-Rao} bound $Var(\hat{\tau}) = 1/(N \mathcal{F}_Q(\tau|\rho_\tau))$. On the other hand, the QFI itself is upper bounded by the variance of the generator of the transformation, i.e, the Hamiltonian H, 
\begin{equation}
    \mathcal{F}_Q(\tau|\rho_\tau) \leq 4~\Delta (H) \leq 1.
    \label{eq:qfibound}
\end{equation}

Here, the variance of the Hamiltonian $\Delta(H) = \langle H^2 \rangle -\langle H \rangle ^2$ and the averages are obtained in the probe state. The upper bound is saturated when the probe is initialized in the optimal state—typically a coherent superposition of the eigenstates of $H$ . 

Against this background, a typical estimation problem (see Fig. \ref{fig:unitaryest}) can be formulated as follows: a probe initialized in the state $\ket{\psi}$ gets encoded with the parameters through the unitary $U$, $\ket{\psi_\mathbb{\lambda}} = e^{-i H \tau} \ket{\psi}$. An appropriate measurement $M$ represented by a positive operator valued measure (POVM) needs to be selected to optimize the FI about the parameter to be estimated. Here, $\bm{\lambda} = (\tau, \theta,\phi)$ corresponds to the set of parameters. In the qubit case, the Hamiltonian $H = \hat{\sigma}.\hat{n}/2$ where $\hat{n} = \sin{\theta}\cos{\phi}~\hat{x} + \sin{\theta}\sin{\phi}~\hat{y} + \cos{\theta}~\hat{z} \equiv (\sin{\theta}\cos{\phi},\sin{\theta}\sin{\phi},\cos{\theta})$ is the (general) rotation axis in spherical polar coordinates, $\hat{\sigma} = (X,Y,Z)$ denotes the vector of Pauli matrices. In the asymptotic limit of many statistical trials ($N \rightarrow \infty$), the 
\begin{equation}
    Var(\hat{\tau}) = 1/(N \mathcal{F}_Q(\tau|\rho_\tau)) = 1/N.
    \label{eq:qsat}
\end{equation}

If the Hamiltonian $H$ is specified, then the saturation mentioned above can be realized. If $H$ is unknown, then optimal estimation of the phase $\tau$ is not possible using the single qubit protocol (see \cite{song2024agnostic}). 

As an illustrative example, consider the probe qubit to be initialized in the $x-z$ plane (without loss of generality). Let the corresponding Bloch vector be given as $\hat{\psi} = \sin{\beta}~\hat{x} + \cos{\beta}~\hat{z}$, where $\beta$ determines the orientation in the plane. Now, if $\beta = \pi/2$ or rather $\hat{\psi} = \hat{x}$, and the measurement is fixed as $M = Y$, then the FI (around $\tau = 0$) is maximum only when $\theta = 0$--i.e., along $\hat{z}$ (or $\pi$ corresponding to -$\hat{z}$).
The choice of the measurement along $\hat{y}$ is that it helps in tracking the direction of the maximal change in the probe state $\hat{\psi}$ as it undergoes the rotation. 

In other words, conventional phase estimation necessitates the knowledge of the Hamiltonian before preparing the appropriate probe state and choosing the ensuing measurement. Absent this knowledge, conventional phase estimation is simply not possible.

Consequently, considering two instants of time $T_1$ and $T_2$ ( with $T_2 > T_1$ according to a chronology respecting (CR) observer) such that $T_1$ corresponds to the preparation of the probe state and $T_2$ corresponds to that of the measurement, the preceding discussion implies that a single qubit based sensor works only if the Hamiltonian is specified before initializing the probe state and measurement. In other words, for a single qubit sensor, to achieve maximum FI, the optimization of the state and measurement, given the Hamiltonian, is the same as optimizing the Hamiltonian for a given probe state and the measurement \cite{song2024agnostic,chapeau2016optimizing}. 

\begin{figure}[!tbp]
  \centering
  \includegraphics[width=0.4\textwidth]{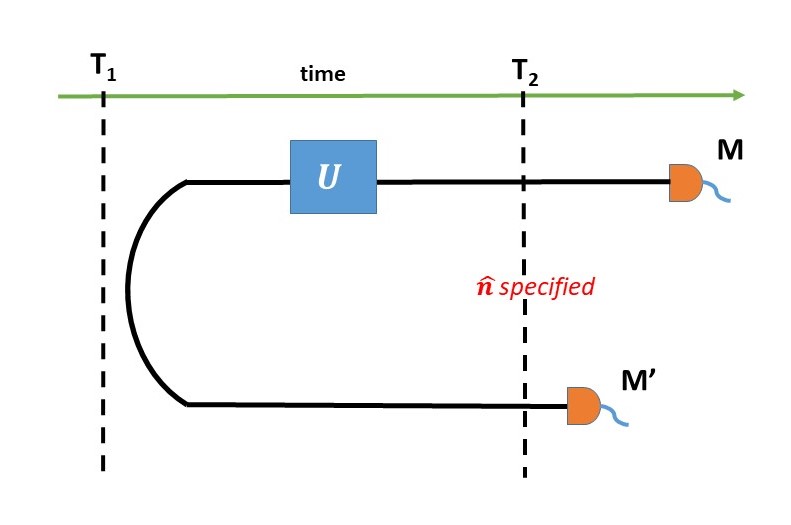}
   \caption{\justifying{Entanglement leveraged hindsight sensing protocol--The $\bigcup$ represents the (initial) entangled state of the probe-ancilla pair. At $T_1$, the Hamiltonian is not revealed. At $T_2$, after the \textit{unknown} transformation $U$ acts on the probe qubit, the Hamiltonian is revealed which will help one to choose the ancilla measurement appropriately. This further helps in interpreting the measurement (M') as the `preparation' of the probe state at $T_2$ which travels back in time, flips its state at $T_1$ and undergoes the transformation $U$ with eventually getting measured later (M).}}
   \label{fig:hindsight}
\end{figure}

\emph{Entanglement leveraged hindsight sensing protocol}: To overcome the limitation of the single-qubit protocol, \cite{song2021quantum,song2024agnostic} explored entanglement as an alternative route to enable phase estimation. Here, we briefly review the hindsight sensing protocol introduced in \cite{song2024agnostic} which employ entanglement in the initial joint state of the probe-ancilla qubit pair.  In this protocol, at time $T_1$ (see Fig. \ref{fig:hindsight}), the entangled state (say singlet) of the probe and the ancilla is prepared. The probe then undergoes the transformation $U$. After this, at time $T_2$, the Hamiltonian responsible for the transformation is revealed. The ancilla is measured appropriately, i.e, in a direction orthogonal to $H$. Using the connection between CTCs and entanglement, the entire sensing protocol is reinterpreted as the time-travel narrative for the probe qubit. To this end, the narrative exploits the connection to effectively prepare a probe state in retrospect to overcome the lack of prior knowledge of the Hamiltonian. The measurement on the ancilla ($M'$) is interpreted as the optimal `preparation' of the probe state in the future (at $T_2$) which then travels backwards in time, flips its state at $T_1$, and then undergoes the transformation $U$ traveling forwards again. A subsequent measurement (M) is carried out later at/after time $T_2$. 

This scenario is illustrated using the typical example of the probe-ancilla pair in an entangled state (say a singlet),
\begin{equation}
    \ket{\Psi^{-}} = \frac{\ket{l}_{P}\ket{\overline{l}}_{A} - \ket{\overline{l}}_{P} \ket{l}_{A}}{\sqrt{2}}.
    \label{eq:singlet}
\end{equation}

Here P and A denotes the probe and ancilla respectively. $\ket{l}$ and $\ket{\overline{l}}$ are the eigenstates of $\hat{\sigma}.\hat{l}$ with $\hat{l}$ being an arbitrary unit vector on the Bloch sphere. After preparing the probe-ancilla pair in the singlet state at time $T_1$, the transformation $U = e^{-i~\tau(\hat{\sigma}.\hat{n})/2}$ is applied on the probe. At time $T_2$, the Hamiltonian ($(\hat{\sigma}.\hat{n})/2$) is revealed as $\sigma_{z}/2$ implying that $\hat{n} = \hat{z}$. This helps in choosing the measurement $M'$ on the ancilla state which in turn helps in the optimal `preparation' of the probe state as mentioned above. Notice that the eigenstates of the optimal measurement operator $M'$ ($\ket{\pm} = (\ket{0} \pm \ket{1})/\sqrt{2}$) correspond to the coherent superposition of the eigenstates of the Hamiltonian ($\propto \sigma_{z}$). Consequently, a measurement $M$ along $\hat{y}$ helps in maximizing the FI ($=1$) for the parameter $\tau$ (around $\tau = 0$). 
This process mirrors the single qubit sensing protocol considering the time-travel narrative for the probe qubit. From the optimal state `preparation' in the future at $T_2$ to traveling backwards in time till $T_1$, when the flip of the state occurs, and then  undergoing the transformation U with a final measurement, this time-travel sensing protocol is exactly similar to the single qubit sensing protocol.
Crucially, by measuring the ancilla ($M'$) after acquiring the knowledge of $\hat{n}$, the FI can be maximized when optimization over the measurement $M$ is performed. 

Alternatively, if one fixes both measurements $M$ and $M'$ a priori, then the protocol constrains $\hat{n}$: the Hamiltonian direction, defined uniquely up to a plane, required to maximize the FI.  Selection of the measurement M' along $\hat{x}$ in the future will then have to eventually lead us to `apply' the Hamiltonian H along $\hat{z}$ in the past for the measurement $M$ chosen along $\hat{y}$ later so that it leads to the maximization of the FI (around $\tau = 0$). In other words, the time-travel aspect of the protocol analyzed for the measurement fixing scenario (the case when $M$ and $M'$ are known and $\hat{n}$ is unknown) is exactly equivalent to the case when the measurement $M'$ on the ancilla is chosen $\hat{n}$-dependently. 

\textit{Coherently controlled superposition (CCS) enabled sensing}:--Previously, in the hindsight sensing protocol, the measurement on the ancilla is chosen based on the knowledge of the applied Hamiltonian. The proposed protocol de-limits this criterion and allows a phase estimation strategy which is irreverent to the knowledge of the Hamiltonian. We term this as the coherently controlled superposition (CCS) sensing strategy. 

As before, consider two time instants, $T_1$ and $T_2$, defined with respect to a CR observer and an unknown unitary operator: $U(\tau) = e^{-i H (\tau)}$. Now, $U^{-1}(\tau) = U^{\dag}(\tau) = e^{i H (\tau)} = e^{-i (-H) (\tau)} = e^{-i H (-\tau)} = U(-~\tau)$, which gives the connotation that if $U$ rotates the probe \textit{anti-clockwise} (`forward' in time) by $\tau$, $U^{\dag}$ is responsible for \textit{clockwise} (`backward' in time) rotation by $-~\tau$ generated by the Hamiltonian `-H' \cite{opc}. To this end, let the action of the two (unknown--the parameters in the Hamiltonian is not known) unitaries $U_0 = U$ and $U_1 = U^{\dag} (=U^{-1})$ on the probe be controlled using the states of an ancilla qubit. If the ancilla is in the state $\ket{0}$, then the probe is acted upon by $U_0$ and if it is in $\ket{1}$, then $U_1$ acts on it. In other words, the state of the ancilla decides which unitary transformation (clockwise or anti-clockwise) does the probe undergo. Moreover, let us consider the combined probe-ancilla pair to be in the initial product state at time $T_1$,

\begin{equation}
\rho_{AP}(0)~=~\ket{q}\bra{q}^{A}\otimes~\rho_{P}(0),
\label{eq:ini}
\end{equation}

where `q' can be either 0 or 1; `A' and `P' denote the ancilla and probe respectively. The total unitary transformation on the combined probe-ancilla pair is given as,
\begin{equation}
L = \ket{0}\bra{0}^{A}\otimes U_0 + \ket{1}\bra{1}^{A}\otimes U_1.
\label{eq:totevol}
\end{equation}
The state of the probe-ancilla pair after the controlled transformation is, 
\begin{eqnarray}
\rho_{AP}(\tau) &=& L \rho_{tot} L^{\dag}\nonumber \\
 &=& \ket{0}\bra{0}^{A} \otimes U_{0}(\tau) \rho_{P}(0) U_{0}^{\dag}(\tau)\nonumber \\
 &+& \ket{1}\bra{1}^{A} \otimes U_{1}(\tau) \rho_{P}(0) U_{1}^{\dag}(\tau)
\label{eq:totstate}
\end{eqnarray}
That is, when the ancilla A is initialized in the state $\ket{q}$, the probe P undergoes the transformation either through the unitary $U_0$ or $U_1$ depending upon the value of q. As such, if A is prepared in an incoherent mixed state, i.e., ($\ket{0}\bra{0}^{A} + \ket{1}\bra{1}^{A})/2$, the probe P passes either through the unitary $U_0$ leading to $U_{0}(\tau) \rho_{P}(0) U_{0}^{\dag}(\tau)$ or $U_1$ leading to $U_{1}(\tau) \rho_{P}(0) U_{1}^{\dag}(\tau)$ with probability $1/2$. On the other hand, when the ancilla is prepared in $\ket{q}_{A}(\alpha) = \cos(\alpha) \ket{0}_{A} + \sin(\alpha) \ket{1}_{A}$ we obtain,

\begin{eqnarray}
\rho_{AP}(\tau;\alpha) = \cos^{2}(\alpha) \ket{0}\bra{0}^{A} \otimes~U_{0}(\tau) \rho_{P}(0) U_{0}^{\dag}(\tau) \nonumber \\
+ \sin^{2}(\alpha) \ket{1}\bra{1}^{A} \otimes~U_{1}(\tau) \rho_{P}(0) U_{1}^{\dag}(\tau) \nonumber \\
+ \cos(\alpha) \sin(\alpha) (\sum_{\substack{q,r = 0,1 \\ q \neq r}} \ket{q}\bra{r}^{A} \otimes ~U_{r}(\tau) \rho_{P}(0) U_{q}^{\dag}(\tau)). 
\label{eq:alphastate}
\end{eqnarray}

Now, for $\alpha = \pi/4$, Eq.\ref{eq:alphastate} can be written as 
\begin{eqnarray}
\rho_{AP}(\tau;\pi/4) = \frac{1}{2}(\ket{0}\bra{0}^{A} \otimes~U_{0}(\tau) \rho_{P}(0) U_{0}^{\dag}(\tau)) \nonumber \\
+ \frac{1}{2}(\ket{1}\bra{1}^{A} \otimes~U_{1}(\tau) \rho_{P}(0) U_{1}^{\dag}(\tau)) \nonumber \\
+\frac{1}{2}(\sum_{\substack{q,r = 0,1 \\ q \neq r}} \ket{q}\bra{r}^{A} \otimes ~U_{r}(\tau) \rho_{P}(0) U_{q}^{\dag}(\tau)). 
\label{eq:+state}
\end{eqnarray}

By (only) measuring the ancilla along $X$, the outcome probabilities are found to be,
 \begin{eqnarray}
     P_+ &=& \cos^2(\tau/2),\nonumber \\
     P_- &=& \sin^2(\tau/2).
    \label{eq:ancprob}
\end{eqnarray}
 The FI about the angle of rotation $\tau$ can now be easily calculated without the knowledge of the Hamiltonian \textit{and that this FI is optimal,i.e, $FI = 1$.} This is a striking and distinctive result since it is the probe qubit which undergoes the phase encoding and the effect of it is witnessed on the ancilla--a clear manifestation of the coherent control mechanism --contrary to conventional phase estimation protocols \cite{giovannetti2006quantum,giovannetti2011advances,chapeau2021quantum}. \\
    
\emph{Hindsight control:--} In \cite{song2024agnostic}, leveraging the entanglement of the probe-ancilla pair (a maximally entangled state), the authors interpreted the metrological protocol using the time-travel narrative for the probe qubit: the optimal probe state is effectively prepared in the future which then travels back in time to undergo the unknown unitary transformation and then eventually goes back to the future to get measured. In our protocol too, the strategy can be interpreted using the time-travel narrative. The main difference being that here it is the ancilla that plays the role of the time-traveler. 

\begin{figure}[ht!]
\centering 
        \includegraphics[width=1.0\linewidth]{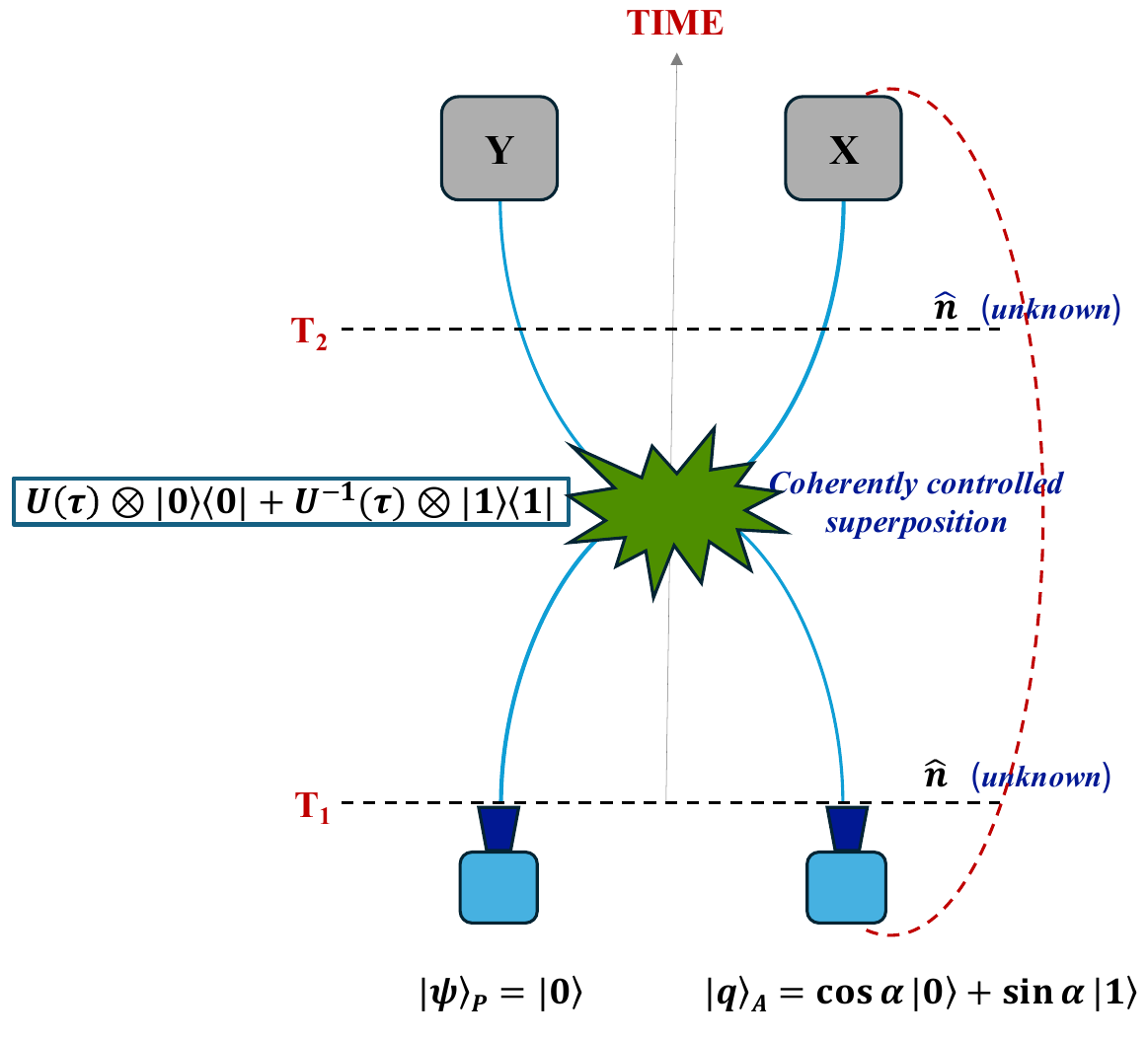}
  \caption{\justifying{The probe and the ancilla qubit start in the state $\ket{0}$ and $\cos{\alpha}\ket{0}+\sin{\alpha} \ket{1}$ respectively. The ancilla qubit interacts with the probe qubit via a controlled superposition (see Eq.\ref{eq:totevol}). The knowledge of the Hamiltonian ($\hat{n}$) is not known before the interaction as well as after the interaction. The probe-ancilla pair is subsequently measured along $\hat{y}$ and $\hat{x}$ similar to the case of Fig.\ref{fig:hindsight}. \textit{Hindsight control}: The dashed red arrow suggests an interpretation of the time-traveling \textit{coherent} ancilla which gets measured in the future at $T_2$, returns back in time at $T_1$ carrying the coherence and turns in as the input state that coherently controls the interaction with the probe qubit and then travels back to the future. The coherence preserving measurement of the ancilla is leveraged for this time-traveling interpretation so that the Hamiltonian-irreverent phase estimation is possible.)}}
   \label{fig:agnostic}
  \end{figure}

The state of the ancilla as a result of the measurement in the coherence preserving $X$ basis in the future at $T_2$ is sent back in time till $T_1$ to start as the initial state controlling the interaction with the probe. This \textit{backward-flowing} information helps the probe to evolve either under $U$ or $U^{-1}$ or a coherent superposition of $U$ and $U^{-1}$ depending on the state of the ancilla (see Eq.\ref{eq:totevol}). This could correspond to the interaction of a chronology respecting probe with a chronology violating ancilla echoing but differing fundamentally from earlier models based on leveraging entanglement \cite{LMG+11,LMGR+11,gfp}. Furthermore, the state of the ancilla acts like a fixed point in the interaction extracting all the information from a transformation (causal) it didn't undergo.

Inspired by these time-traveling scenarios, we describe the present metrological protocol in terms of a \textit{coherence leveraged time-travel} narrative: the probe qubit interacts with a time-traveling coherent ancilla from the future and undergoes the coherently controlled superposition of $U$ and $U^{-1}$. The coherent ancilla \cite{ico} thus enables Hamiltonian-agnostic phase estimation by effectively undergoing an optimal coherent control state "preparation" from the future (see Fig.\ref{fig:agnostic}). 

\begin{figure}[!ht]
  \centering
  \includegraphics[width=0.25\textwidth]{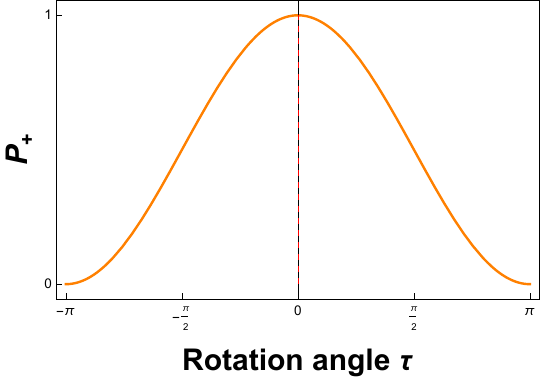}
  \caption{\justifying{The success probability of the outcome $+1$ for the measurement X on the ancilla. This denotes the success probability of simulating the time-travel of the ancilla exhibiting its role as the \textit{coherent} hindsight control.}}
  \label{fig:probvst}
   \end{figure}

Coincidentally, the measurement probabilities $P_{\pm}$ (see Eq.\ref{eq:ancprob}) corresponds to the success probabilities of finding the ancilla in the coherent state $\ket\pm$ (Fig.\ref{fig:probvst} shows the variation of $P_+$ w.r.t $\tau$). It is this coherence preserving measurement (aligned with the initial state of the ancilla)--that reflects as an unique  interference of $U$ and $U^{-1}$ resulting on the probe--which gives rise to the non-classical response on the ancilla's (measurement) outcome statistics. \emph{Furthermore, it is pertinent to note that, similar to the agnostic protocol of \cite{song2024agnostic}, in the CCS sensing protocol too, neither post-selection nor any data discarding is performed}.

\emph{Hindsight sensing vs CCS sensing}: The hindsight and CCS protocols consider measurements on both the probe and ancilla separately. However, the main difference between them corresponds to the unique way the phase gets encoded via the unknown unitary on the probe. In the Hindsight sensing, the unknown unitary acts on the probe after it has been entangled with the ancilla. In contrast, the CCS protocol begin with an initially separate ancilla and probe that jointly undergo a coherently controlled interaction where in the unknown transformation on the probe is applied \textit{conditioned} on the state of the ancilla. Depending on the state of the ancilla, either the probe transforms under $U$ or $U^{-1}$ or a quantumly possible unique \textit{superposition} together. Specifically, the unknown unitary $$U(\tau,\theta,\phi) = e^{-i  \tau~\hat{\sigma}.\hat{n}(\theta,\phi)/2},~\text{with}~~ \bm{\lambda}:= (\tau,\theta,\phi),$$ which are the parameters to be estimated in the unitary. Among these three unknown parameters, we are interested in estimating the angle of rotation $\tau$ alone independent of the axis parameters $(\theta,\phi)$. 

To asses this, we compute the FI matrix using the probabilities from the measurement $Y \otimes X$ on the probe (initially in the state $\ket{0}$) and the controlling ancilla ($\ket{+}$) respectively and find it in the block-diagonal form 
\begin{equation}
F(\bm{\lambda}) = \begin{pmatrix}
    &1 &0 &0\\
     &0 &F_{\theta\theta} &F_{\theta\phi}\\ 
      &0 &F_{\phi\theta} &F_{\phi\phi}
\end{pmatrix} ~\text{for all}~ \bm{\lambda}.
\label{eq:mat}
\end{equation}

This confirms that the CCS protocol achieves agnostic phase estimation as the FI about $\tau$ ($F_{\tau\tau} = 1$) is independent to the variation of the axis parameters ($\theta, \phi$): the terms $F_{i\tau} = 0 = F_{\tau i}$ for $i \in \{\theta, \phi\}$. Note that the terms $F_{ij}$ reflects the information about the correlation between the parameters $i,j$ and, if non-zero, signals that the estimation of one parameter changes due to variation in the other.

\textit{In contrast, the FI matrix corresponding to the hindsight sensing differs from the block-diagonal form of Eq.(\ref{eq:mat}); instead, it is a full matrix with non-zero off-diagonal terms $F_{i\tau}, F_{\tau i}$ for $i \in \{\theta, \phi\}$. This indicates that the hindsight sensing mandates the knowledge of the axis. In other words, the estimation of the phase is not agnostic with respect to the Hamiltonian in the hindsight sensing protocol.}

\emph{Conclusion:--} We have introduced a novel metrology strategy for estimating an unknown phase, capable of achieving optimal Fisher information, without requiring prior knowledge of the Hamiltonian generating it. Unlike existing approaches that rely heavily on entanglement-assisted methods--using entangled state and entangling measurement--our approach leverages quantum coherence for achieving axis-agnostic phase estimation. This simplicity offers a resource-efficient route towards Hamiltonian-agnostic phase estimation.

To this end, our results point towards some relevant research directions for future investigations. It will be interesting to extend the CCS sensing protocol to multi-parameter estimation or estimation using higher-dimensional systems which may reveal new coherence-based metrological advantages--with potential applications in quantum sensing and metrology. On the experimental side, our protocol can be easily implemented in platforms where entanglement is an expensive resource such as photonic \cite{GLT+24, SSQ+24} or superconducting quantum systems \cite{friis2015coherent}. In such platforms, coherent control can be naturally realized which can offer practical cost-effective demonstration of Hamiltonian-agnostic sensing. On the quantum foundational side, investigations on the connection between metrology, time-symmetry \cite{ico} and information flow can be explored. In addition, an extension of the above protocol considering relativistic scenarios for estimating parameters (say, proper times/accelerations) will be considered as one of the lines of our future investigations. 

\emph{Acknowledgments.--} 
I thank Jorge Escand{\'o}n for discussions  during the early stage of this work. I'm especially indebted to A. R. Usha Devi and Akshata Shenoy H for extensive (and at times, exhausting) discussions during the entire stage of the work. I thank Marco Barbieri, Matteo Paris for their comments during the Quantum 2025,18-24th May 2025, INRiM, Turin, Italy. Also, I'm grateful to David R. M. Arvidsson-Shukur for  constructive comments/suggestions on an earlier version of this manuscript. HSK thanks NCN Poland, ChistEra-2023/05/Y/ST2/00005 under the project Modern Device Independent Cryptography (MoDIC).
\bibliography{reff}

\begin{thebibliography}{50}
\bibitem [{gfp()}]{gfp}%
  \BibitemOpen
  \href@noop {} {}\bibinfo {note} {In the well-known grandfather paradox, the actions of the time-traveling grand-daugther who interacts with her grandfather could have an effect on her own existence. Such logical conundrums are avoided either by framing self-consitency conditions \cite{Novi83} (for quantum theory), chronology protection conjecture \cite{Haw92} or probabilistic interpretations \cite{LMGR+11}. Our metrological protocol echoes the spirit of these themes by offering a probabilistic interpretation in terms of the coherence preserving measurement of the ancilla.}\BibitemShut {Stop}%
\end{thebibliography}

\section{Appendix}
\subsection{A: A short note on (quantum) parameter estimation}

A typical estimation problem (see Fig. \ref{fig:unitaryest} in the main text) can be stated as follows: a probe initialized in the state $\ket{\psi}$ gets encoded with the parameters through the unitary $U$, $\ket{\psi_\mathbb{\lambda}} = U_\lambda \ket{\psi}$. An appropriate measurement corresponding to a positive operator valued measure (POVM) M is selected and measured. From the resulting outcome probability distribution one optimizes the Fisher information (FI) about the parameter being estimated. Here, 
$\bm{\lambda} := (\lambda_1,\lambda_2,\lambda_3)$ corresponds to the set of parameters. In the case of a qubit, the encoding unitary $U = e^{-i H(\theta,\phi) \tau}$ where the Hamiltonian $H(\theta,\phi) = \hat{\sigma}.\hat{n}(\theta,\phi)/2$ where $\hat{n}(\theta,\phi) = \sin{\theta}\cos{\phi}~\hat{x} + \sin{\theta}\sin{\phi}~\hat{y} + \cos{\theta}~\hat{z} \equiv (\sin{\theta}\cos{\phi},\sin{\theta}\sin{\phi},\cos{\theta})$ is the (general) rotation axis in spherical polar coordinates, $\hat{\sigma} = (X,Y,Z)$ denotes the Pauli vector and $\bm{\lambda} \equiv (\tau,\theta,\phi)$ are the set of unknown parameters. 

Notably, there are two strategies to optimize the sensing protocol (i.e, to find the optimal FI): one is by finding the appropriate probe state and use it to calculate the quantum FI (QFI) via Eq.\ref{eq:qfifull} as shown below or by finding an optimal measurement M so that the classical FI saturates the QFI. Given that the probability distribution $p(m|\bm{\lambda})$ captures the outcomes $m$ from the measurement of M conditioned on the parameters,  the classical FI matrix is defined as;
\begin{equation}
   F_{M,ij}(\bm{\lambda}) =  \sum_{m} p(m|\bm{\lambda})[\partial_i \log p(m|\bm{\lambda})][\partial_j \log p(m|\bm{\lambda})],
    \label{eq:cfifull}
\end{equation}
where the partial derivatives $\partial_i\equiv \partial/\partial\lambda_i$, i.e, the partial derivative with respect to the parameters.
An \textit{estimator} $\hat{\bm{\lambda}}$ is constructed which is used to estimate the values of the parameters from the outcomes of the measurement M. A property of the estimator which reflects the randomness in the outcomes of the POVM M corresponds to the estimator being termed as \textit{unbiased} which essentially means that the expectation value of the estimator is equal to the parameter being estimated, $\mathbb{E}(\hat{\bm{\lambda}}) = \sum_m p(m|\bm{\lambda})~ \hat{\bm{\lambda}} = \bm{\lambda}$. 

On the other hand, the QFI (for a pure state) is defined as a matrix (QFIM) with coefficients as 
\begin{eqnarray}
   \mathcal{F}_{Q,ij}(\bm{\lambda}|\rho_{\bm{\lambda}}) &=&  4(\braket{\partial_i \psi(\bm{\lambda})}{\partial_j \psi(\bm{\lambda})} \nonumber \\
   &-& \braket{\partial_i \psi(\bm{\lambda})}{\psi(\bm{\lambda})} \braket{\psi(\bm{\lambda})}{\partial_j \psi(\bm{\lambda})}),
    \label{eq:qfifull}
\end{eqnarray}
with $i,j \in \{\tau,\theta,\phi\}$. 

Incidentally, the \textit{ Cram\'er-Rao} bound is expressed as \cite{rao1992information,cramer1999mathematical},
\begin{equation}
    Cov(\hat{\bm{\lambda}}) \geq \frac{1}{N} F(\bm{\lambda})^{-1} \geq \frac{1}{N} \mathcal{F}_Q(\bm{\lambda}|\rho_{\bm{\lambda}})^{-1},
    \label{eq:qcr}
\end{equation}
where $Cov(\hat{\bm{\lambda}})$ is the covariance matrix, the elements of which quantify the interdependence amongst the parameters. The second inequality follows from the fact that for any measurement, $F(\bm{\lambda}) \leq \mathcal{F}_Q(\bm{\lambda}|\rho_{\bm{\lambda}}).$ The elements of the covariance matrix are given by  $Cov(\hat{\bm{\lambda}})_{ij} = \mathbb{E}(\lambda_i \lambda_j) - \mathbb{E}(\lambda_i) \mathbb{E}(\lambda_j)$.

Now, if a probe state is pure and we are interested in estimating only one parameter alone, say the rotation angle (phase), the QFI is computed using,
\begin{equation}
   \mathcal{F}_Q(\tau|\ket{\psi_\tau}) =  4(|\braket{\partial_\tau \psi(\tau)}|^2 - |\braket{\partial_\tau \psi(\tau)}{\psi(\tau)}|^2)
    \label{eq:qfi}
\end{equation}
The classical FI (see the main text) is defined as 
\begin{equation}
    F_M(\tau):=\sum_{m}\frac{(\partial_{\tau}p(m|\tau)^2}{p(m|\tau)}.
    \label{cfi}
\end{equation}

Employing the best measurement along with the limit of many statistical trials ($N \rightarrow \infty$), $F(\tau)$ becomes equal to $\mathcal{F}_Q(\tau|\rho_\tau)$ leading to the saturation of the CR bound:  
\begin{equation}
    Var(\hat{\tau}) = 1/(N \mathcal{F}_Q(\tau|\rho_\tau)) = 1/N.
    \label{eq:qsat1}
\end{equation}
Note: If the state is not pure, the QFI is defined as: $\mathcal{F}_Q(\lambda_i|\rho_{\lambda_i})) = {\rm Tr}[L_{\lambda_i}~\partial_{\lambda_i} \rho_{\lambda_i}]$ in terms of the \textit{symmetric logarithmic derivative} $L$ satisfying $\partial_{\lambda_i} \rho_{\lambda_i} = L_{\lambda_i}~\rho_{\lambda_i} +\rho_{\lambda_i}~L_{\lambda_i}$. The elements of the QFI matrix is then given as:$\mathcal{F}_Q(\bm{\lambda}|\rho_{\bm{\lambda}})_{i,j} := \rm{Tr}[L_{\lambda_i} \partial_{\lambda_j}~\rho_{\bm{\lambda}}]$  \cite{H69,braunstein1994statistical,fujiwara1995quantum}.

Coming back to the case of estimating the three parameters $\bm{\lambda}$ we require the covariance of the estimator $\hat{\bm{\lambda}}$ to be as less as possible giving rise to a performance indicator known as \textit{mean squared error} (MSE) of the estimator $\hat{\bm{\lambda}}$. To this end, it is defined as 
\begin{equation}
   MSE[\hat{\bm{\lambda}}]=\text{Tr}(Cov(\hat{\bm{\lambda}})).
    \label{mse}
\end{equation}

Eq.\ref{eq:qcr} suggests us a cue through which we can escape from actually constructing an optimal estimator. To this end, there are two ways out: (1) construct a reasonably good probe state and calculate QFI, (2) Use this QFI and find an optimal measurement M from which the closest (saturating) classical FI can be computed.  
Given this scenario, it becomes important to choose an appropriate measurement M from which the estimation of all the parameters becomes optimal. However, contrary to our expectation, if it is easier to find a measurement which optimizes the FI about $\lambda_i$, it might turn out to be not optimal for $\lambda_j$ due to the underlying incompatibility of the parameters. As such, to remain irreverent to the performance of the two estimators $\hat{\bm{\lambda_1}}$ and $\hat{\bm{\lambda_1}}$, Eq.\ref{eq:qcr} is modified into a scalar form through the introduction of a \emph{weight matrix} W leading to,

\begin{equation}
\resizebox{\columnwidth}{!}{$
   \text{Tr}[W\, \text{Cov}(\hat{\bm{\lambda}})] 
   \geq \frac{1}{N} \text{Tr}[W F(\bm{\lambda})^{-1}] 
   \geq \frac{1}{N} \text{Tr}[W \mathcal{F}_Q(\bm{\lambda}|\rho_{\bm{\lambda}})^{-1}]
$}
\label{eq:hqcr}
\end{equation}

For the case at hand, we can set the weight matrix as \cite{goldberg2021intrinsic,song2024agnostic},
\begin{equation}
W = \begin{pmatrix}
    &1 &0 &0\\
     &0 &0 &0\\
      &0 &0 &0
\end{pmatrix},
\end{equation} implying that we are only interested in the estimation of a single parameter, that is, the rotation angle $\tau$ alone.  

Now, if we find the QFI matrix as ,
\begin{equation}
\mathcal{F}_{Q}(\bm{\lambda|\rho_{\bm{\lambda}}}) = \begin{pmatrix}
    &\mathcal{F}_{\tau\tau} &0 &0\\
     &0 &\mathcal{F}_{\theta\theta} &\mathcal{F}_{\theta\phi}\\ 
      &0 &\mathcal{F}_{\phi\theta} &\mathcal{F}_{\phi\phi}
\end{pmatrix} ~\text{for all}~ \bm{\lambda},
\label{eq:mat1}
\end{equation}
then, the estimation of the angle of rotation $\tau$ can be performed agnostic to the knowledge of the other parameters $\theta$, $\phi$. In other words, $Var(\hat{\bm{\tau}}) \geq 1/(N \mathcal{F}_{\tau\tau}) = 1/(N \mathcal{F}_Q(\tau|\rho_\tau))$ no matter what the lower right hand $2\times2$ block is. 
Alternatively, if we calculate the classical FI matrix and if it is given by Eq.\ref{eq:mat}, then it corresponds to agnostic phase estimation. 

\subsection{B: Calculation of the Fisher information matrix and quantum coherence as a resource}

Consider the initial state of the ancilla and the probe state to be in
the pure product form $\rho_{AP}(0)~=~\ket{+}\bra{+}^{A}\otimes~\rho_{P}(0)$, where $\rho_{P}(0)$ ($= \ket{0}\bra{0}$) is the state of the probe. Evolve the state using $L = \ket{0}\bra{0}^{A}\otimes U + \ket{1}\bra{1}^{A}\otimes U^{-1}$ after which measure the ancilla and the probe along X and Y respectively. The outcome probabilities are given as: 
\begin{widetext}
\begin{align}
    P_{AP}(+1,+1) &= Tr[(\ket{+}\bra{+}^A \otimes \ket{+_y}\bra{+_y}^P)~\rho_{AP}(\tau;\pi/4)] = \frac{\cos^2(\tau/2)}{2}, \nonumber
    \\
    P_{AP}(+1,-1) &= Tr[(\ket{+}\bra{+}^A \otimes \ket{-_y}\bra{-_y}^P)~\rho_{AP}(\tau;\pi/4)] = \frac{\sin^2(\tau/2)}{2}(1+\sin{2\theta} \sin{\phi}),\nonumber
    \\
     P_{AP}(-1,+1) &= Tr[(\ket{-}\bra{-}^A \otimes \ket{+_y}\bra{+_y}^P)~\rho_{AP}(\tau;\pi/4)] = \frac{\cos^2(\tau/2)}{2},\nonumber
     \\
     P_{AP}(-1,-1) &= Tr[(\ket{-}\bra{-}^A \otimes \ket{-_y}\bra{-_y}^P)~\rho_{AP}(\tau;\pi/4)] = \frac{\sin^2(\tau/2)}{2}(1-\sin{2\theta} \sin{\phi}).
    \label{eq:fullprob}
\end{align}
\end{widetext}

From these probabilities, the FI matrix can be computed: 

\begin{equation}
\resizebox{\columnwidth}{!}{$
F^{\ket{0}}_{\bm{XY,\lambda}} = \begin{bmatrix}
    1 & 0 & 0\\
    0 &-(\frac{(4 \cos^{2}(2 \theta) \sin^{2}(\frac{\tau}{2}) \sin^{2}(\phi)}{-1 + \sin^{2}(2 \theta)~\sin^{2}(\phi)}) &-(\frac{(\sin(4 \theta) \sin^{2}(\frac{\tau}{2}) \sin(2\phi)}{-2 + 2\sin^{2}(2 \theta)~\sin^{2}(\phi)})\\ 
      0 &-(\frac{(\sin(4 \theta) \sin^{2}(\frac{\tau}{2}) \sin(2\phi)}{-2 + 2\sin^{2}(2 \theta)~\sin^{2}(\phi)}) &-(\frac{(\sin^{2}(2 \theta) \sin^{2}(\frac{\tau}{2}) \cos^{2}(\phi)}{-1 + \sin^{2}(2 \theta)~\sin^{2}(\phi)})
\end{bmatrix}$},
\label{eq:fimxz}
\end{equation}

 which is, of the form given in Eq.\ref{eq:mat1}. The subscript $XY$ (and $\ket{0}$) on the l.h.s of the FI matrix signals the measurement used on the ancilla and the probe (initial state of the probe) respectively. The FI about the angle of rotation $\tau$ is maximum ($=1$) and is estimated being agnostic to the knowledge of the axis of rotation (Hamiltonian). Though the FI matrix is singular, however, the saturation of the QFI for the estimation of $\tau$ is independent of this singular nature. 

 On the other hand, if the initial state of the probe is maximally mixed, then 
\begin{equation}
 F^{\frac{\mathds{1}}{2}}_{\bm{XM,\lambda}}= \begin{pmatrix}
    &1 &0 &0\\
     &0 &0 &0\\
      &0 &0 &0
\end{pmatrix},
\end{equation}
where $M$ can be any measurement on the Bloch sphere. The $X$ measurement on the ancilla, though, gives the outcome probabilities (as given in Eq.\ref{eq:ancprob} in the main text) from which the estimation on $\tau$ is achieved agnostic to the knowledge of the axis-parameters. An important point to note here is that, when the probe is maximally mixed, no entanglement is generated dynamically. However, agnostic sensing of the phase is still achieved. \textit{This possibility arises from the use of quantum coherence in the ancilla qubit to create coherent superpositions of the unitary transformations. This enables interference between alternative evolutions of the probe, even when the probe carries no coherence itself.} Such a resource  — \textit{coherent} control of operations — has no classical analog \cite{abbott2020communication, friis2015coherent,SSQ+24,GLT+24}.

In the case of an arbitrary pure ancilla initialized in the state $\cos{\alpha} \ket{0} + \sin{\alpha} \ket{1}$ where $\alpha$ captures the coherence --as in the $l_1$-norm of coherence \cite{baumgratz2014quantifying} ($C(\alpha) = 2 |(\rho_{anc}(\alpha))_{01}| = \sin{2 \alpha}$), then, coherence acts as a quantifiable resource, which then suggests that the classical (and quantum) FI  behaves exactly as a monotone of this coherence \cite{feng2017quantifying}. The classical FI can be computed using the measurement statistics of the ancilla as $ F_\tau ^{Coh}(C(\alpha)) = \frac{C(\alpha) ^2 \sin ^2(\tau )}{1-C(\alpha) ^2 \cos ^2(\tau )}$ ($C(\alpha) \longrightarrow f$ in Eq. \ref{eq:noisyfic} below). On the other hand, the QFI can be computed directly from the joint state, $\rho_{AP}(\tau;\alpha)$ (see Eq. \ref{eq:alphastate} in the main text), as $C(\alpha)^2 + (1 - C(\alpha)^2)\sin^{2} (2 \theta )$. Notice that the saturation of the classical FI with the QFI happens \emph{only} at $\alpha = \pi/4$ leading to agnostic phase estimation. 
\begin{figure}[!tbp]
  \centering
  \subfloat[]{\includegraphics[width=0.4\textwidth]{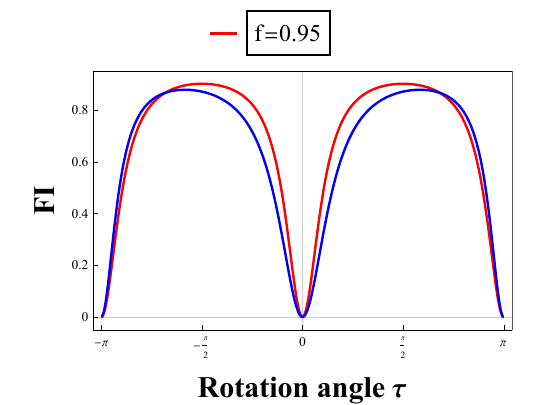}}
  \\
  \subfloat[]{\includegraphics[width=0.23\textwidth]{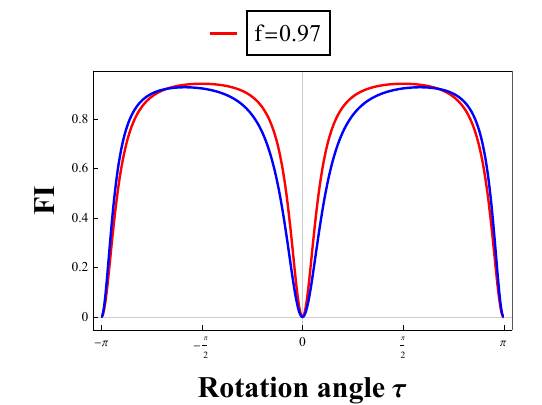}}
   \subfloat[]{\includegraphics[width=0.25\textwidth]{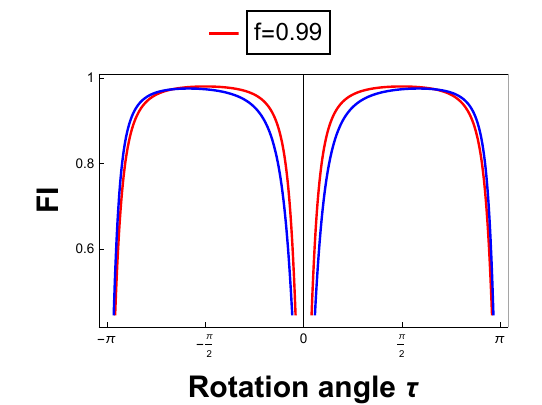}}
    \caption{\justifying{(a) Plot showing FI (about $\tau$) vs the angle of rotation $\tau$. The red curve corresponds to the CCS sensing case (noisy ancilla) and the blue one to the agnostic metrology of \cite{song2024agnostic} (noisy preparation of the singlet). Here, the noise strength $f = 0.95$.
    (b) The noise strength $f = 0.97$.
    (c) The noise strength $f = 0.99$ Notice that in all the three cases, the CCS sensing strategy performs slightly better than that involving entangled state and measurement.}}
    \label{fig:compfi}
\end{figure}

\subsection{C: Comparing the noise robustness of the sensing protocols}

Considering a realistic implementation of the coherent superposition implies that the control qubit will suffer from imperfections either in the preparation or at the measurement stage, which directly degrades its ability to sustain the superposition essential to our protocol. To this end, we model this imperfection as a white noise (depolarizing channel) acting on the ancilla immediately before the joint interaction. Specifically, we replace the ideal ancilla $\ket{+}\bra{+}$ state $\longrightarrow f \ket{+}\bra{+} +\frac{1}{2} (1-f ) \mathds{1}$ where $f \in [0,1]$ quantifies strength of the noise. $f = 1$ implies no effect of noise on the state and $f = 0$ means maximum decoherence (and that the $\ket{+}$ is reduced to a completely mixed state $\mathds{1}/2$).

After this channel noise, the probabilities of measuring the ancilla in the $X$ basis become,
\begin{eqnarray}
    P_+(\tau;f) &=& \frac{1}{2} (1 + f  \cos (\tau ))\\
    P_-(\tau;f) &=& \frac{1}{2} (1 - f  \cos (\tau )).
    \label{eq:noisyprob}
\end{eqnarray}
 With this, the FI evaluates to
 \begin{equation}
     F_\tau ^{Coh}(f) = \frac{f ^2 \sin ^2(\tau )}{1-f ^2 \cos ^2(\tau )}.
     \label{eq:noisyfic}
 \end{equation}
 We can compare this FI to that reported in \cite{song2024agnostic} for the case of the entangled state and entangling measurement,
 \begin{equation}
 \resizebox{\columnwidth}{!}{$
     F_\tau^{Ent} (f) = -\frac{(1-4 f)^2 \sin ^2(\tau )}{((4 f-1) \cos (\tau )+2 f-5) ((4 f-1) \cos (\tau )+2 f+1)}$}.
     \label{eq:noisyfie}
 \end{equation}

 Notice that the plots in Fig. \ref{fig:compfi} clearly indicate that the introduction of noise in the ancilla (entangled state) starts limiting the maximal FI. Eqns \ref{eq:noisyfic},\ref{eq:noisyfie} reflect that the advantage of agnostic phase estimation due to coherence (entanglement)  gets severely limited due to the presence of noise. However, it is important to highlight that under such noisy considerations, our phase estimation protocol performs \emph{slightly better} than its entanglement leveraged counterpart \cite{song2024agnostic}.  
 
\end{document}